\documentclass[a4paper,reqno,12pt]{amsart}
\usepackage[all]{xy}           
\usepackage{amssymb}           
\usepackage[utf8]{inputenc}
\usepackage{hyperref}
\usepackage{eucal}
\usepackage{epsfig}
\usepackage{pst-grad} 
\usepackage{pst-plot} 
\numberwithin{equation}{section}

\newtheorem{definition}{Definition}[section]
\newtheorem{lemma}[definition]{Lemma}
\newtheorem{theorem}[definition]{Theorem}
\newtheorem{proposition}[definition]{Proposition}

\newtheorem{remarkth}[definition]{Remark}

\renewcommand{\emph}[1]{{\bfseries\itshape{#1}}}

\makeatletter
\newcommand\prol{\@ifstar{\@proldf}{\@prolpf}}  
\def\@prolpf{\@ifnextchar[{\@prolpf@wrt}{\@prolpf@}}
\def\@prolpf@wrt[#1]#2{\@ifnextchar[{\@prolpf@wrt@at{#1}{#2}}{\@prolpf@wrt@{#1}{#2}}}
\def\@prolpf@wrt@at#1#2[#3]{\prolsymbol^{#1}_{#3}#2}
\def\@prolpf@wrt@#1#2{\prolsymbol^{#1}#2}
\def\@prolpf@#1{\@ifnextchar[{\@prolpf@at{#1}}{\@prolpf@@{#1}}}
\def\@prolpf@at#1[#2]{\prolsymbol_{#2}#1}
\def\@prolpf@@#1{\prolsymbol#1}
\def\@proldf{\@ifnextchar[{\@proldf@wrt}{\@proldf@}}
\def\@proldf@wrt[#1]#2{\@ifnextchar[{\@proldf@wrt@at{#1}{#2}}{\@proldf@wrt@{#1}{#2}}}
\def\@proldf@wrt@at#1#2[#3]{\prolsymbol^{*#1}_{#3}#2}
\def\@proldf@wrt@#1#2{\prolsymbol^{*#1}#2}
\def\@proldf@#1{\@ifnextchar[{\@proldf@at{#1}}{\@proldf@@{#1}}}
\def\@proldf@at#1[#2]{\prolsymbol^*_{#2}#1}
\def\@proldf@@#1{\prolsymbol^*#1}
\def\prolsymbol{\mathcal{T}}
\makeatother

\begin{document}

\title[Notes on Lagrangian continuum mechanics]{Notes on Lagrangian continuum mechanics}

\author[V. M. Jim\'enez]{V\'ictor Manuel Jim\'enez}
\address{V\'ictor Manuel Jim\'enez:
Universidad Nacional de Educación a Distancia (UNED), \\
Departamento de Matemáticas Fundamentales. \\
Calle de Juan del Rosal 10, 28040, Madrid, Spain} \email{victor.jimenez@mat.uned.es}

\thanks{Author acknowledge financial support from the Spanish Ministry of Science and Innovation, under grants Grant PID2019-106715GB-C21 and PID2022-137909NB-C2 and the Severo Ochoa Programme for Centres of Excellence in R$\&$D” (CEX2019-000904-S). Author also acknowledge the finantial support from MIU and European Union-\textit{NextGenerationUE}.}
 \subjclass[2000]{}

\begin{abstract}
In Continuum Mechanic a simple material body $\mathcal{B}$ is represeted by a three-dimensional differentiable manifold and the configuration space is given by the space of embeddings $Emb \left( \mathcal{B} , \mathbb{R}^{n} \right)$. We use the topology of infinite-dimensional manifold of this space, to present the first variation formula for Lagrangian mechanics.
\end{abstract}

\maketitle

\tableofcontents



\section{Manifold of differentiable maps. Embeddings}

Let $M$ and $N$ be two (finite dimensional) differentiable manifolds. We will begin with a brief introduction to a topology of differentiable manifold which can constructed onto the manifold of differentiable maps. Along this paper the finite dimensional manifolds will be considered as manifolds without boundary, second contable and Hausdorff.\\
First, we will define the following topologies onto the space of continuous functions:
\begin{itemize}
\item[(i)] \textbf{Compact-open (or $CO$) topology:} A subbasis for this topology is given by the family of sets of the way,
$$ \{ f \in \mathcal{C}^{0} \left( M ,N \right) \ : \ f \left( K \right) \subseteq U \},$$
where $K$ is a compact subset of $M$ and $U$ is an open subset of $N$.
\item[(ii)] \textbf{Wholly open (or $WO$) topology:} A basis for this topology is given by the family of sets of the way,
$$ \{ f \in \mathcal{C}^{0} \left( M ,N \right) \ : \ f \left( M \right) \subseteq U \},$$
where $U$ is an open subset of $N$.
\item[(iii)] \textbf{The graph (or $WO^{0}$) topology:} We define the following map:
$$ \Gamma : \mathcal{C}^{0} \left(M,N \right) \rightarrow \mathcal{C}^{0} \left(M,M \times N \right),$$
where, for each $f \in \mathcal{C}^{0} \left(M,N \right)$, $\Gamma \left( f \right)$ is the graph of $f$. So, the $WO^{0}-$topology is the initial topology associated to this map with the $WO-$topology in $\mathcal{C}^{0} \left(M,M \times N \right)$. Then, a basis for this topology is given by the family of sets of the way,
$$ \{ f \in \mathcal{C}^{0} \left( M ,N \right) \ : \ \Gamma \left( f \right) \left( M \right) \subseteq U \},$$
where $U$ is an open subset of $M \times N$.
\end{itemize}

Note that, the $WO-$topology is not a Hausdorff topology because surjective maps cannot be separated. On the other hand, the $CO-$topology is a Hausdorff topology. Finally, the $WO^{0}-$topology is finer than the compact-open topology and, hence, is a Hausdorff topology.\\
Next, we want to use these topologies to induce topologies into the spaces of $\mathcal{C}^{k}-$differentiable maps for $k \geq 1$. In order to do this, we will introduce the space of $k-$jets.\\
Let $f: U \subseteq M \rightarrow V \subseteq N$ be a $\mathcal{C}^{k}-$differentiable local function for some $0 \leq k \leq \infty$. The $k-$jet of $f$ at the point $p \in U$ is given an equivalence class containing those functions with the same value and first $k$ derivatives at $p$ and it will be denoted by $j^{k}_{p} f$. We shall specify the equivalence relation using coordinates (the particular choice of coordinate system does not matter). The space of $k-$jets from $M$ to $N$ is denoted by $J^{k} \left( M , N \right)$. For $k=0$,
$$ J^{0} \left( M,N \right) \triangleq M \times N.$$
Suppose $k < \infty$. Let $\left( x^{i} \right)$ and $\left( y^{j} \right)$ be two local coordinate systems on open sets $U\subseteq M$ and $V \subseteq N$ respectively. Then, we can construct a local system of coordinates over $J^{k} \left( M,N \right)$ as follows:
\begin{equation}\label{1}
J^{k}\left(U,V\right) : \left(x^{i} , y^{j}, y^{j}_{i_{1}}, y^{j}_{i_{1},i_{2}}, \dots, y^{j}_{i_{1}, \dots, i_{k}} \right),
\end{equation}
where, for each $j^{k}_{p} f $,
\begin{itemize}
\item $x^{i} \left(j^{k}_{p} f\right) = x^{i} \left(p\right)$
\item $y^{j} \left(j^{k}_{p} f\right) = y^{j} \left( f \left( p \right)\right)$
\item $y^{j}_{i_{1} , \dots , i_{l}}\left(j^{k}_{p} f\right)  = \dfrac{\partial^{l} \left(y^{j}\circ l\right)}{\partial x^{i_{1}} \dots \partial x^{i_{l}}_{| p} }, \ 1 \leq l \leq k$
\end{itemize}
So, for $0 \leq k < \infty$, $J^{k} \left( M , N \right)$ is a (finite dimensional) manifold.\\
On the other hand, suppose $k = \infty$. Notice that $J^{\infty} \left( M , N \right)$ can be seen as the projective limits in the category of Hausdorff topological spaces of spaces $J^{k} \left( M , N \right)$. Obviously, the space of $\infty-$jets is not a finite dimensional manifold. But it can be proved the following statement:
\begin{lemma}
Let $M,N$ be (finite dimensional) manifolds. Then $J^{\infty} \left( M , N \right)$ is a second contable metrizable manifold.
\end{lemma}
Next, for each $k-$differentiable function $f: M \rightarrow N$ we can define the \textit{$k-$prolongation of $f$} as follows,
$$ j^{k}f : M \rightarrow J^{k} \left( M , N \right),$$
such that, for all $p \in M$
$$ j^{k}f \left( p \right) = j^{k}_{p} f.$$
Thus, we define the map
\begin{equation}\label{2}
j^{k} : \mathcal{C}^{k} \left( M , N \right) \rightarrow \mathcal{C}^{0} \left( M , J^{k} \left( M , N \right) \right).
\end{equation}
In this way, we define the $CO^{k}-$\textit{topology} (resp. \textit{$WO^{k}-$topology}) onto $\mathcal{C}^{k} \left( M , N \right)$ as the initial topology associated to $j^{k}$ with the $CO-$topology (resp. $WO-$topology) in $\mathcal{C}^{0} \left( M , J^{k} \left( M , N \right) \right)$. Notice that for $k=0$ the $WO^{0}-$topology coincide with the defined above. Furthermore, it is easy to prove that all these topologies are Hausdorff.\\
Then, we can claim the following:
\begin{itemize}
\item[(i)] The $CO^{k}-$toplogy on $\mathcal{C}^{k} \left( M , N \right)$ is completely metrizable, thus a Baire-topology.
\item[(ii)] The $WO^{k}-$topology on $\mathcal{C}^{k} \left( M , N \right)$ is, again, a Baire-topology. Each $CO^{k}$-closed subset of $\mathcal{C}^{k} \left( M , N \right)$ is a Baire space too in $WO^{k}$.
\item[(iii)] A basis of the $WO^{\infty}-$topology for $\mathcal{C}^{\infty} \left( M , N \right)$ is given by the sets,
$$\{ g \in \mathcal{C}^{\infty} \left( M , N \right)  \ : \ j^{k}g \left( M \right) \subseteq U \}, \ k = 0, 1 , 2, \dots $$
where $U$ is an open subset of $N$.
\end{itemize}
Next, let be a finite dimensional vector bundle $\pi : A \rightarrow X$ and $\Gamma \left( A \right)$ the space of sections of $\pi$.\\
Let us notice that any topology in $\mathcal{C}^{\infty} \left(  X , A \right)$ induces a topology onto $\Gamma \left( A \right)$ as the induced topology. So, we will choose the $WO^{\infty}-$topology onto $\Gamma \left( A \right)$. However, this is not a topological vector space.
\begin{proposition}
Let $\Gamma_{c} \left( A \right)$ be the space of sections with compact support equipped with the induced topology of $\Gamma \left( A \right)$ ($WO^{\infty}-$topology). Then, $\Gamma_{c} \left( A \right)$ is the maximal topological vector space contained in $\Gamma \left( A \right)$.
\end{proposition}
So, these will be our modelling spaces over the space of differentiable maps.\\

Fix $f \in \mathcal{C}^{\infty} \left( M,N \right)$. We want to define a local chart of $f$. First, we may construct the pullback vector bundle $f^{*}TN$, i.e., the pullback which satisfies that the diagram

\vspace{0.5cm}
\begin{picture}(375,50)(50,40)
\put(180,20){\makebox(0,0){$M$}}
\put(250,25){$f$}               \put(210,20){\vector(1,0){80}}
\put(310,20){\makebox(0,0){$N$}}
\put(160,50){$pr_{1}$}                  \put(180,70){\vector(0,-1){40}}
\put(320,50){$\pi_{N}$}                  \put(310,70){\vector(0,-1){40}}
\put(160,80){\makebox(0,0){$f^{*}TN \subseteq M \times TN$}}
\put(250,85){$pr_{2}$}               \put(210,80){\vector(1,0){80}}
\put(310,80){\makebox(0,0){$TN$}}
\end{picture}

\vspace{35pt}

is commutative. We will identify $f^{*}TN$ with the union of the tangent spaces of $N$ at the points in the image of $f$.\\

Now, choose $g$ an auxiliar Riemannian metric on $N$. Let $exp_{U} : U \subseteq TN \rightarrow N$ the exponential map of $g$ over an neighbourhood of the zero section of $TN$ such that the map $\left( \pi_{N} , exp_{U} \right) : U \rightarrow V \subseteq N \times N$ is a diffeomorphism, where $V$ is an open neighbourhood of the diagonal.\\
Next, for $f \in \mathcal{C}^{\infty} \left( M,N \right)$ consider the following subset of $\mathcal{C}^{\infty} \left( M,N \right)$ ,
$$ U_{f} := \{ g \in \mathcal{C}^{\infty} \left( M,N \right) \ : \ \left( f,g \right) \left( M \right) \subseteq V, \  f \sim g \},$$
where $f\sim g$ if there exists a compact subset $K \subset M$ such that
$$f_{|K^{c}} = g_{|K^{c}},$$
where $K^{c}$ is the complementary of $K$ in $M$. So, we will construct the following map,
$$
\begin{array}{rccl}
\Psi_{U_{f}} : & U_{f} & \rightarrow & \Gamma_{c}\left( f^{*}TN \right) \\
& g  &\mapsto & \Psi_{U_{f}} \left( g \right)\\
\end{array}
,$$
such that,
$$\Psi_{U_{f}} \left( g \right) \left( x \right) = \left( x ,\left( \pi_{N} , exp_{U} \right)^{-1} \left(f\left( x\right), g \left( x \right)\right)\right), \ \forall x \in M.$$
This map is bijective onto the set
$$V_{f} := \{ \alpha \in \Gamma_{c}\left( f^{*}TN \right) \ : \ \alpha \left( M \right) \subseteq f^{*}U \},$$
with $f^{*}U \triangleq f^{*}TN \cap U$. This set is, obviously, an open in $\Gamma_{c} \left(  f^{*}TN \right)$ (with the $WO^{\infty}-$topology).\\
The inverse map is given by
$$ \Psi_{U_{f}}^{-1} \left( \alpha \right) = exp \circ \alpha,
 \ \forall \alpha \in \Gamma_{c}\left( f^{*}TN \right).$$
We can consider in $\Gamma_{c} \left( f^{*}TN \right)$ the induced topology from the $WO^{\infty}-$topology in $\mathcal{C}^{\infty} \left( M , f^{*}TN \right)$. This topology turns $\Gamma_{c} \left( f^{*}TN \right)$ into a bornological locally convex vector space. However, we will consider a finer topology onto $\Gamma_{c} \left( f^{*}TN \right)$ than the $WO^{\infty}-$topology which is given by the final topology associated to the family of differentiable curves onto $\Gamma_{c} \left( f^{*}TN \right)$. This topology is called $c^{\infty}-$topology in \cite{AKPWM}.\\
It turns out that the family $\{ \left( U_{f} , \Psi_{U_{f}} \right) \}_{f}$ over $\mathcal{C}^{\infty} \left( M,N \right)$ defines an atlas which does not depend on the metric $g$ and, hence, $\mathcal{C}^{\infty} \left( M,N \right)$ is an infinite dimensional manifold.\\

\begin{theorem}\label{3}
Let $M$ and $N$ be smooth finite dimensional manifolds. Then the space $\mathcal{C}^{\infty} \left( M,N \right)$ of all smooth mappings from $M$ to $N$ is a smooth manifold, modeled on spaces $\Gamma_{c}\left( f^{*}TN \right)$ of smooth sections with compact support of pullback bundles along $f$. This structure has separable connected components, is smoothly paracompact and Lindelöf.\\
If $M$ and $N$ are connected, the space of embeddings $Emb \left( M , N \right)$ is an open of $\mathcal{C}^{\infty} \left( M,N \right)$.
\end{theorem}
To study a detailed construction of this differentiable structure see (\cite{AKPWM}).\\
\begin{remarkth}
\rm
We could construct another differentiable structure onto $\mathcal{C}^{\infty} \left( M , N \right)$ refining the $WO^{\infty}-$topology of this space including the classes of each map as opens in the new topology. So, the same atlas gives us another structure of differentiable manifold. This construction can be seen in \cite{PWM} (see also \cite{HIR}).\\
On the other hand, suppose that $M$ is compact. Then, the same charts gives us a differentiable structure onto $\mathcal{C}^{k} \left( M,N \right)$ with the $WO^{k}-$topology for all $k \leq \infty$. The modelled spaces are spaces of $k-$differentiable sections of a vector bundle over $M$ with the same topology. Then,
\begin{itemize}
\item[(i)] For $k < \infty$, $\mathcal{C}^{k} \left( M,N \right)$ is an infinite dimesional manifold modelled over Banach spaces.\\
\item[(ii)] For $k = \infty$, $\mathcal{C}^{\infty} \left( M,N \right)$ is an infinite dimesional manifold modelled over Fréchet spaces.\\
\end{itemize}
These topologies can be seen developed in detail in \cite{BINZ}.
\end{remarkth}

\section{Lagrangian mechanics}
In this section we will introduce a geometrical formulation of continuum mechanics. First, let us present the picture for the Lagrangian formulation for classical mechanics. A \textit{point of mass} $P$ is zero-dimensional manifold which can be completely described by its mass and spatial position. A \textit{configuration of $P$} is a position of a point mass and it is denoted $q \in \mathbb{R}^{d}$. For systems formed by $N$ point masses, the configuration is a vector $\left( q_{1}, \dots , q_{N} \right) \in \mathbb{R}^{dN}$ given by the position vectors of each of the point masses. The set of all possible configurations of a system of $N$ point masses is called its \textit{configuration space}. In the absence of constraints, the configuration space is $\mathbb{R}^{dN}$ (or some open subset of $\mathbb{R}^{dN}$). In general, the configuration space is assumed to be a finite-dimensional manifold $Q$. Notice that any finite dimesional manifold can be considered as an embedded submanifold of $\mathbb{R}^{n}$.\\
A \textit{classical mechanical system} in the sense of Lagrange consists of
\begin{itemize}
\item[(1)] A configuration space $Q$.
\item[(2)] A scalar function $L: TQ \rightarrow \mathbb{R}$
\end{itemize}
The Lagrangian function is \textit{decomposable} if it can be written as
\begin{equation}\label{15}
L = T + V
\end{equation}
where $T$ is given by a Riemannian metric on $Q$ and $V = dv : TQ \rightarrow \mathbb{R}$, where $v: Q \rightarrow \mathbb{R}$. $T$ and $V$ are called, respectively the \textit{kinetic} and \textit{potential energies} of the system.\\
The kinetic energy $T$ represents the inertial properties of the system, that is, its tendency to remain in its state of rest or motion (as declared by Newton's first law). The potential energy $V$ has to do with the applied forces that want to disturb this state (as declared by Newton's second law).\\
Now, we will come back to continuum mechanics. A \textit{body} is a $m-$dimensional manifold $\mathcal{B}$. The \textit{space of the body $\mathcal{B}$} is a $\mathbb{R}^{n}$, for some $m \leq n$.  A \textit{configuration} of $\mathcal{B}$ is an embedding $\phi : \mathcal{B} \rightarrow \mathbb{R}^{n}$. Hence, assuming that the body does not have contraints, the configuration space will be
$$  Emb \left( \mathcal{B} , \mathbb{R}^{n} \right).$$
We endow this space with the topology of differentiable manifold given in Theorem \ref{3}. Other authors choose work with the space of \linebreak$k-$differentiable embeddings. This is because, as we have noticed, this manifold is modelled over Banach spaces (see for example \cite{RSE}, \cite{RKU}).\\
With this, it is natural to define a \textit{configuration space} as an embedded submanifold
$$ \mathcal{Q} \leq Emb \left( \mathcal{B} , \mathbb{R}^{n} \right) .$$
Notice that, as particular family of configurations we have the emdedding space,
$$  Emb \left( \mathcal{B} , \mathcal{S} \right),$$
where $\mathcal{S}$ is a manifold of dimension $n$.\\
A \textit{Lagrangian function} will be a differentiable function
$$ \mathcal{L} : T \mathcal{Q} \rightarrow \mathbb{R}.$$
The Lagrangian function is said to be \textit{descomposable} if it can be written as
\begin{equation}\label{4}
\mathcal{L} = T - V,
\end{equation}
where $T$ is a weak Riemannian metric on $\mathcal{Q}$ and $V = dv$ where $v : \mathcal{Q} \rightarrow \mathbb{R}$. $T$ and $V$ are called the \textit{kinetic} and \textit{potencial} energies of the system respectively.\\
A \textit{force} $\alpha_{\phi}$ at a configuration $\phi$ is given by an element of the cotangent bundle $T^{*}_{\phi}\mathcal{Q}$. A \textit{force field} is an $1-$form,
$$ \alpha :\mathcal{Q} \rightarrow T^{*} \mathcal{Q}.$$
A force field $\alpha$ is said to be conservative or to derive from a potential if $\alpha$ is exact. That is, a force field is conservative if there exist a differentiable map $f : \mathcal{Q} \rightarrow \mathbb{R}$ such that
$$ \alpha = df.$$
The objective of Mechanics is to predict, out of given initial conditions $\phi_{0}$ and $v_{\phi_{0}}$ at time $t = 0$, the system trajectory for times $t > 0$ within a certain time interval. We now formulate a fundamental axiom, which we call \textit{Lagrange's postulate}.\\\\

\textbf{Postulate 1} \textit{For a decomposable mechanical system, in the absence of any force field, the system follows a geodesic of the Levi-Civita connection induced by the kinetic-energy}.\\\\

Let $\overline{g} : T\mathcal{Q} \times_{\mathcal{Q}} T \mathcal{Q} \rightarrow \mathbb{R}$ be a weak Riemannian metric. Then, the associated map
$$ \overline{g}^{b} : T \mathcal{Q} \rightarrow T^{*} \mathcal{Q},$$
is not a diffeomorphism. In fact, we can only claim that this map is injective. Therefore, we cannot ensure the existence of the Levi-Civita coonnection.\\
Next, we will present a metric over the space of configurations which has a Levi-Civita connection.\\
Let $g$ be a Riemannian metric in $\mathcal{S}$. Then, we can induce a weak Riemannian metric onto $\mathcal{Q}$ in the following way:
\begin{equation}\label{10.2}
\overline{g} \left( \phi \right) \left( s_{1} , s_{2} \right) = \int_{\mathcal{B}} [g \left( s_{1} , s_{2} \right) \circ \phi ]Vol \left( \phi^{*} g \right),
\end{equation}
where $\phi \in \mathcal{Q}$ and $ s_{1} , s_{2} \in \Gamma_{c} \left( \phi^{*}T \mathcal{S} \right)$.\\
Now, we will give an sketch of the construction of the Levi-Civita connection. 

\begin{definition}
\rm
Let $M$ be a (possibly infinite dimensional) manifold. A \textit{connector} in $M$ is a differentiable map $\kappa : T^{2} M \rightarrow TM$ such that
\begin{itemize}
\item[(i)] It is a retraction of the vertical lift $V: TM \rightarrow T^{2} M$.
\item[(ii)] It is linear respect to the vector bundle structures $\pi_{T M}$ and $T \pi_{M}$.
\end{itemize}
\end{definition}
Let $\left( U , \varphi \right)$ be a local chart on $M$. Then, locally, a connector $\kappa$ is characterized by a map,
$$ \kappa: U \times E \times E \times E \rightarrow U \times E,$$
such that
$$ \kappa \left( x , u , v , w \right) = \left( x , w- \Gamma \left( x, u,v \right) \right),$$
where $\Gamma$ is bilinear in the last two coordinates. The map $\Gamma$ is called the \textit{Christoffel symbol of $\kappa$}.\\
Notice that, any connector $\kappa$ induces a covariant derivative $\nabla$ on the manifold $M$ by the following identity
\begin{equation}\label{5}
\nabla_{X}Y = \kappa \circ TY \circ X.
\end{equation}
Then, the map $\Gamma$ is, indeed, the Christoffel symbol of $\nabla$ over $U$.\\
It is easy to prove that, in the finite dimensional case, any covariant derivative induces a unique connector which satisfies Eq. (\ref{5}).\\

Taking into account this fact, let us consider the Levi-Civita connection of the Riemannian metric $g$ and $\kappa^{0}$ its associated connector. Then, the induced map (by composition) ${\kappa^{0}}^{*}$ of $\kappa^{0}$ over $\mathcal{Q}$ is a connector on $\mathcal{Q}$. Thus, we can construct a covariant derivative on $\mathcal{Q}$ with associated connector ${\kappa^{0}}^{*}$ by Eq. (\ref{5}). We will denote this covariant derivative by $\nabla^{0}$.\\
Now, using the Koszul formula, the covariant derivative associated $\nabla$ to $\overline{g}$ (if there exists) satisfies that
\begin{scriptsize}
\begin{eqnarray*}
\overline{g} \left( X, \nabla_{Y}Z - \nabla^{0}_{Y} Z \right) &=& \dfrac{1}{2} \{ - \left( \nabla^{0}_{X} g \right) \left( Y , Z \right) + \left( \nabla^{0}_{Y} g \right) \left( Z , X \right) + \left( \nabla^{0}_{Z} g \right) \left( X,Y \right)\\
&-& g \left( Tor^{0} \left( X,Y \right) , Z \right) - g \left( Tor^{0} \left( Y,Z \right) , X \right) + g \left( Tor^{0} \left( Z,X \right) , Y \right) \},
\end{eqnarray*}
\end{scriptsize}
where $Tor^{0}$ is the torsion of the covariant derivative $\nabla^{0}$.\\
We can prove that the torsion of this map is zero but $\nabla^{0} g$ is not zero. Hence, $\nabla^{0}$ is not the Levi-Civita connection. However we can prove that, for all $X, Y , Z \in \frak X \left( \mathcal{Q} \right)$, the right-side of the above equation is in $g^{b} \left( \nabla_{Y}Z - \nabla^{0}_{Y} Z \right)$. So, by injectivity, there exists a unique covariant derivative $\nabla$ which is the Levi-Civita connection of $\overline{g}$. For details see \cite{KAI}.\\
Let $\gamma : I \rightarrow \mathcal{Q}$ be a differentiable curve. Then, for any $V $ vector field along $\gamma$, we can define de derivative of $V$ along $\gamma$ as follows:
$$V^{'} \triangleq \{ \nabla_{\dot{\gamma} \left( t \right)} V \} \left( t \right),$$
where $\dot{\gamma}$ is the \textit{velocity of the curve}, i.e.,
$$ \dot{\gamma} \left( t \right) = T_{t}\gamma \left( \dfrac{\partial}{\partial t_{|t}} \right), \ \forall t \in I$$
So, locally, have that
\begin{equation}\label{12}
V^{'} = \dfrac{\partial V}{\partial t_{|t}} - \Gamma \left( \gamma ; V , \dot{\gamma} \right).
\end{equation}
where $\Gamma$ is the Christoffel symbol of $\nabla$ over $U$. The \textit{acceleration of $\gamma$} is given by the curve
$$ \nabla_{\dot{\gamma}} \dot{\gamma} \triangleq \ddot{\gamma} \left( t \right).$$
Thus, the postulate 1 can be reformulated as follows:\\

\textbf{Postulate 1} \textit{In the absence of any force field, a decomposable system follows a trajectory of vanishing acceleration.}\\\\

Consider now the case of the presence of a non-vanishing force field $\alpha$. According to the Newtonian viewpoint, we postulate that the trajectory of the system is governed by the second-order ODE system
\begin{equation}\label{6}
\overline{g}^{b} \left( \ddot{\gamma} \left( t \right) \right) = \alpha \circ \gamma
\end{equation}
This equation is called the \textit{equation of motion}. Notice that this equation only makes sense if $\alpha \in \overline{g}^{b} \left( T \mathcal{Q} \right)$, these kind of $1-$forms are called \textit{admissible}. Therefore, in this case the equation of motion cannot be expressed for any force field (as in the finite dimensional case). In fact, notice that for all $\phi \in Emb \left( \mathcal{B} , \mathcal{S} \right)$ the restriction of $g^{b}$ to $\phi^{*} T \mathcal{S}$
$$ g^{b}_{\phi} : \phi^{*} T \mathcal{S} \rightarrow \phi^{*} T^{*} \mathcal{S},$$
is a isomorphism. Hence, it induces an isomorphism of topological vector spaces
$$ g^{b}_{\phi} : \Gamma_{c} \left( \phi^{*} T \mathcal{S}\right) \rightarrow  \Gamma_{c} \left(\phi^{*} T^{*} \mathcal{S}\right).$$
So, let be a force $\Lambda_{\phi} \in T_{\phi}^{*} \mathcal{Q}$. Then, $\Lambda_{\phi} \in \overline{g}^{b} \left( T_{\phi} \mathcal{Q} \right)$ if and only if there exists $\sigma_{\phi} \in \Gamma_{c} \left(\phi^{*} T^{*} \mathcal{S}\right)$ such that
\begin{equation}\label{14}
\Lambda_{\phi} \left( s \right) = \int_{\mathcal{B}} \sigma_{\phi} \left( s \right)Vol \left( \phi^{*} g \right).
\end{equation}
So, roughly speaking, any force in $\mathcal{Q}$ is admissible if and only if it is given by a integral of a force field in $\mathcal{S}$. Finally, Eq. (\ref{6}) can be equivalently written in the following way
\begin{equation}\label{7}
\ddot{\gamma} \left( t \right)  = X \circ \gamma
\end{equation}
where $X = {\overline{g}^{b}}^{-1} \left( \alpha \right) \in \frak X \left( \mathcal{Q} \right)$. Note that, Eq. (\ref{7}) makes sense for any vector field on $\mathcal{Q}$. This will be our equation of motion from now on.\\
Finally, we will give another way of working the equation of motion.\\
Let $\gamma : [a,b] \rightarrow \mathcal{Q}$ be a piecewise differentiable curve on $\mathcal{Q}$. Then, the \textit{energy of $\gamma$} is given by
\begin{equation}\label{8}
E \left( \gamma \right) = \dfrac{1}{2} \int_{a}^{b} \overline{g} \left( \dot{\gamma} , \dot{\gamma} \right) dt,
\end{equation}
i.e.,
\begin{equation}\label{10}
E \left( \gamma \right)  = \dfrac{1}{2} \int_{a}^{b} \int_{\mathcal{B}} g \left( \dot{\gamma} , \dot{\gamma} \right)Vol \left( \gamma^{*} g \right) dt
\end{equation}
A \textit{variation of $\gamma$} is a continuous map $\overline{\gamma}: [a,b] \times \left( -\delta , \delta \right) \rightarrow \mathcal{Q}$ such that
\begin{itemize}
\item[(i)] $ \overline{\gamma} \left( t,0 \right) = \gamma \left( t \right), \ \forall t \in [ a,b]$.
\item[(ii)] There exists a finite family of number $\{t_{i} \}_{i=1}^{k} \subseteq [a,b]$ such that $a = t_{0} < t_{1} < \dots < t_{k-1} < t_{k}=b$ and the restriction of $\overline{\gamma}$ to $[t_{i}, t_{i+1}] \times \left(  -\delta , \delta \right)$ is differentiable for all $i$.
\end{itemize}
A variation $\overline{\gamma}$ is said to be \textit{proper} if it is constant over the boundary of $[a,b]$.\\

\begin{definition}
\rm
Let $\overline{\gamma}$ be a variation of a piecewise differentiable curve $\gamma : [a,b] \rightarrow \mathcal{Q}$. Then, we define the \textit{variational vector field of $\overline{\gamma}$} as follows
$$ V \left( t \right) = \dfrac{\partial \overline{\gamma}}{\partial s_{|\left( t,0 \right)}},$$
where $s$ represents the second coordinate.
\end{definition}
Notice that the variational vector field is not a vector field along $\gamma$ in the usual sense because it is not differentiable but piecewise differentiable. Furthermore, for all piecewise vector field $V$ along $\gamma$, there exists a variation of $\gamma$ such that $V$ is the variational vector field associated to this variation. Obviously, if $V$ is (completely) differentiable the associated variation is differentiable too. Finally, it easy to prove that a variation $\overline{\gamma}$ is proper if and only if its variational vector field $V$ is zero in the boundary of the interval.\\

\begin{proposition}[\textit{First variation formula}]
Let $\overline{\gamma}$ be a variation of a piecewise differentiable curve $\gamma : [a,b] \rightarrow \mathcal{Q}$. Then,
\begin{eqnarray*}
\dfrac{\partial E \circ \overline{\gamma}}{\partial s_{|0}} = &-& \int_{a}^{b} \overline{g} \left( V , \ddot{\gamma} \right) dt \\
&-& \sum_{i=1}^{k} \overline{g} \left( V \left( t_{i} \right), \dfrac{\partial \gamma}{\partial t_{t_{i}^{+}}} - \dfrac{\partial \gamma}{\partial t_{t_{i}^{-}}} \right) \\
&-& \overline{g} \left( V \left( a \right) , \dot{\gamma} \left( a \right) \right)  + \overline{g} \left( V \left( b \right) , \dot{\gamma} \left( b \right) \right),
\end{eqnarray*}
where $V$ is the variational vector field of $\overline{\gamma}$ and $\dfrac{\partial \gamma}{\partial t_{t_{i}^{+}}}$ (resp. $\dfrac{\partial \gamma}{\partial t_{t_{i}^{-}}}$) is the right-derivative (resp. left-derivative) of $\gamma$ at $t_{i}$.
\end{proposition}
Now, fix $X \in \frak X \left( \mathcal{Q} \right)$. Let $\gamma : [ a , b ] \rightarrow \mathcal{Q}$ be a solution of the equation of motion (\ref{7}). Then, for all proper differentiable variation $\overline{\gamma}$ of $\gamma$ we have that,
\begin{equation}\label{9}
\dfrac{\partial E \circ \overline{\gamma}}{\partial s_{|0}} = - \int_{a}^{b} \overline{g} \left( V , X \right) dt,
\end{equation}
where $V$ is the variational vector field of $\overline{\gamma}$.\\
Conversely, suppose that Eq. (\ref{9}) is satisfied for all proper differentiable variation $\overline{\gamma}$ of $\gamma$. Let us consider the vector field along $\gamma$,
$$ V \left( t \right) \triangleq \lambda \left( t \right) \left( X \left( \gamma \left( t \right) \right) - \ddot{\gamma} \right) \triangleq  \lambda \left( t \right) Y \left( t \right),$$
where $\lambda \left( a \right) = \lambda \left( b \right) = 0$. Let $\overline{\gamma}$ be a differentiable proper variation of $\gamma$ such that $V$ is its variational vector field.\\
Then, using Eq. (\ref{9}), we have that
\begin{eqnarray*}
0 &=& \int_{a}^{b} \lambda \left( t \right) \overline{g} \left( Y \left( t \right) , Y \left( t \right) \right) dt\\
&=&\int_{a}^{b}\int_{\mathcal{B}} \lambda \left( t ,x \right) g \left( Y \left( t,x \right) , Y \left( t,x \right) \right) Vol \left( \gamma^{*} g \right) dt
\end{eqnarray*}
for all diferentiable function $\lambda$ which satisfies that $\lambda \left( a \right) = \lambda \left( b \right) = 0$. Equivalently, 
$$ \ddot{\gamma} = X \left( t \right).$$
So, Eq. (\ref{9}) for proper differentiable variations gives us a new way of dealing with the equation of motion.\\

\section{Example}
In this section, we will give an explicit form of the equation of motion for a particular example. In order to do this, we will use the Eq. (\ref{9})\\
Suppose that $\mathcal{B}$ and $\mathcal{S}$ are open intervals in $\mathbb{R}$. Then, the metric in the space $\mathcal{Q}$ given by Eq. (\ref{10.2}) (induced by the usual metric in $\mathbb{R}$) is as follows,
\begin{equation}
\overline{g} \left( \phi \right) \left( h,k \right) = \int_{\mathcal{B}} h \left( x \right) k \left( x \right) \mid \phi^{'} \left( x \right) \mid dx,
\end{equation}
for all $\phi \in  \mathcal{Q}$ and $h,k \in \mathcal{C}_{c}^{\infty} \left( \mathcal{B} , \mathbb{R} \right)$. Let $\gamma : [a,b] \rightarrow \mathcal{Q}$ be a differentiable curve and $\overline{\gamma}$ be a proper differentiable variation of $\gamma$. For the sake of simplicity, we will denote the derivative at $t$ (resp. $s$ and $x$) of $\overline{\gamma}$ by $\overline{\gamma}_{,t}$ (resp. $\overline{\gamma}_{,s}$ and $\overline{\gamma}_{,x}$). We will also suppose that $\gamma_{,x} >0$.\\
First, note that
\begin{eqnarray*}
\dfrac{\partial E \circ \overline{\gamma}}{\partial s_{|0}} &=& \dfrac{1}{2}\dfrac{\partial}{\partial_{s_{|0}}} \int_{a}^{b}\int_{\mathcal{B}}\overline{\gamma}_{,t}^{2}\overline{\gamma}_{,x} dxdt \\
&=& \dfrac{1}{2} \int_{a}^{b}\int_{\mathcal{B}} [2\overline{\gamma}_{,ts}\gamma_{,t}\gamma_{,x} + \gamma_{,t}^{2} \overline{\gamma}_{,xs}]dxdt.
\end{eqnarray*}
Notice that
\begin{equation}\label{11}
\int_{a}^{b} \int_{\mathcal{B}}[ \left( \gamma_{,t}\overline{\gamma}_{,s} \gamma_{,x} \right)_{,t} + \dfrac{1}{2} \left( \gamma_{,t}^{2} \overline{\gamma}_{,s} \right)_{,x}]dxdt .
\end{equation}

Without lose of generality we could suppose that $\overline{\gamma}$ (by continuous extension over $\overline{\mathcal{B}}$) is constant over $\partial \mathcal{B}$ (with respect to $s$). Then, we have that Eq. (\ref{11}) is equal to zero.\\
Taking into account this, we have that
\begin{eqnarray*}
\dfrac{\partial E \circ \overline{\gamma}}{\partial s_{|0}} &=& -  \dfrac{1}{2} \int_{a}^{b}\int_{\mathcal{B}} [ 2\gamma_{,tt}\overline{\gamma}_{,s}\gamma_{,x} + 4\overline{\gamma}_{,s}\gamma_{,t}\gamma_{,xt}] \\
&=& - \int_{a}^{b}\int_{\mathcal{B}} [ \gamma_{,tt} + 2\dfrac{\gamma_{,t}\gamma_{,xt}}{\gamma_{,x}}]V \left( t \right) \gamma_{,x}dxdt
\end{eqnarray*}
So, using Eq. (\ref{9}), $\gamma$ is a solution of the equation of motion if and only if
$$0 =  \int_{a}^{b}\int_{\mathcal{B}} [ \gamma_{,tt} + 2\dfrac{\gamma_{,t}\gamma_{,xt}}{\gamma_{,x}} -2X \left( \gamma\left(t \right)\right)]V \left( t \right) \gamma_{,x}dxdt.$$
Therefore, the equation of motion is the following
\begin{equation}
\gamma_{,tt} =  2 [X \left( \gamma\left(t \right)\right) - \dfrac{\gamma_{,t}\gamma_{,xt}}{\gamma_{,x}}]
\end{equation}
with the initial conditions: $\gamma\left( 0 \right) \in Emb \left( \mathcal{B} , \mathcal{S} \right)$, $\gamma_{,t}\left( 0 \right) \in \mathcal{C}_{c}^{\infty} \left( \mathcal{B} , \mathbb{R} \right)$.\\

\begin{remarkth}
\rm
Note that $\gamma \left( t \right) \in Emb \left( \mathcal{B} , \mathcal{S} \right)$ for all $t \in [a,b]$. Then, $\gamma_{,x} $ is not zero at any point. Hence, $\gamma_{,x} >0$ for all $\left( t,x \right) \in [a,b] \times \mathcal{B}$ or $\gamma_{,x} <0$ for all $\left( t,x \right) \in [a,b] \times \mathcal{B}$. So, we can do the above process for the case of $\gamma_{,x} <0$ to get the same equation.
\end{remarkth}

Notice that, using Eq. (\ref{12}), for each $\gamma : [ a , b ] \rightarrow \mathcal{Q}$ geodesic it satisfies that the Christoffel symbols
$$ \Gamma \left( \gamma , \gamma_{,t} , \gamma_{,t} \right) =  \dfrac{\gamma_{,t}\gamma_{,xt}}{\gamma_{,x}}.$$
Finally, taking into account that the torsion is free que have that
\begin{equation}\label{13}
\Gamma \left( \phi , h , k \right) =  -\dfrac{hk_{,x}+ kh_{,x}}{\phi_{,x}}.
\end{equation}
With this, we get an explicit expression of the Levi-Civita conection.

\end{document}